\documentclass[]{spie}  

\usepackage{amsmath,amsfonts,amssymb}
\usepackage{graphicx}
\usepackage[colorlinks=true, allcolors=blue]{hyperref}

\title{Measurements and analysis of optical crosstalk in a microwave kinetic inductance detector array}

\author[a,b]{L. Bisigello}
\author[b]{S.J.C. Yates}
\author[b]{L. Ferrari}
\author[c]{J.J.A Baselmans}
\author[a,b]{A.M. Baryshev}

\affil[a]{Kapteyn Astronomical Institute, University of Groningen, P.O. Box 800, 9700 AV, Groningen, The Netherlands.}
\affil[b]{SRON Netherlands Institute for Space Research Groningen, 9747 AD, Groningen, The Netherlands.}
\affil[c]{SRON Netherlands Institute for Space Research Utrecht, 3584 CA, Utrecht, The Netherlands}

\begin{document} 
\maketitle

\begin{abstract}
The main advantage of Microwave Kinetic Inductance Detector arrays (MKID) is their multiplexing capability, which allows for building cameras with a large number of pixels and good sensitivity, particularly suitable to perform large blank galaxy surveys. However, to have as many pixels as possible it is necessary to arrange detectors close in readout frequency. Consequently KIDs overlap in frequency and are coupled to each other producing crosstalk. Because crosstalk can be only minimised by improving the array design, in this work we aim to correct for this effect a posteriori.
We analysed a MKID array consisting of 880 KIDs with readout frequencies at 4-8 GHz. We measured the beam patterns for every detector in the array and described the response of each detector by using a two-dimensional Gaussian fit. Then, we identified detectors affected by crosstalk above -30 dB level from the maximum and removed the signal of the crosstalking detectors. Moreover, we modelled the crosstalk level for each KID as a function of the readout frequency separation starting from the assumption that the transmission of a KID is a Lorenztian function in power. We were able to describe the general crosstalk level of the array and the crosstalk of each KID within 5 dB, so enabling the design of future arrays with the crosstalk as a design criterion.
In this work, we demonstrate that it is possible to process MKID images a posteriori to decrease the crosstalk effect, subtracting the response of each coupled KID from the original map.

\end{abstract}


\section{INTRODUCTION}
\label{sec:intro}  
In astronomy, several blank imaging surveys at different wavelengths have been carried out to study the formation and evolution of galaxies at different cosmic epochs \cite{Weiss2009,Geach2013}. A multi-wavelength approach is essential to have a more complete view of galaxy properties and, in particular, sub-millimetre observations are necessary to explore the dust component of galaxies. Microwave kinetic inductance detectors (MKID\cite{Day2003,Baselmans2012,Mazin2009}) are the ideal technology to built fast and large cameras, such as A-MKID\footnote{http://www3.mpifr-bonn.mpg.de/div/submmtech/bolometer/A-MKID/a-mkidmain.html} or NIKA\cite{Monfardini2010,Monfardini2011}, to carry out deep and large blank galaxy surveys in the sub-millimetre regime. The main advantage of this technology is the possibility to read out all detectors simultaneously throughout a single readout line. This is possible because each detector is tuned to a specific resonance frequency and they are read out by sending wave tones though the readout line. \par
We used an array of 880 twin-slot antenna coupled hybrid MKIDs made for development and test in view of the SPACEKIDS\footnote{http://www.spacekids.eu}  project. This technology has been already applied to similar array showing good efficiency and sensitivity\cite{Janssen2013}. 
KIDs are tuned to absorb 350 GHz and have resonance frequencies between 4 GHz and 8 GHz with a design separation in readout frequency of 2.64-5.28 MHz and designed quality factors (Q-factors) around 40000.
Detectors are organised in the array such that the nearest spatial neighbours are always separated by at least one other KID in readout frequency domain \cite{Yates2014}, in order to minimise the number of crosstalking KIDs.  The only crosstalk in this array is therefore from the overlapping resonant dips of the KIDs themselves. This is both due by design, to maximise number of KIDs per readout line, and due to scatter in the KID placement due to lithographical and film thickness variations.
 \par
The aim of this paper is to correct for the crosstalk a posteriori, both by describing the point spread function (PSF), as well as by deriving a theoretical model to predict the crosstalk as a function of the separation in readout frequencies of the KIDs from the resonance frequency and the quality factor of each KID. \par

\section{Beam map measurement and PSF characterisation}
\label{sec:beammap}
 \subsection{Measurements}
 In order to analyse the level of crosstalk, we used two types of measurements: beam maps and frequency sweeps. \par
\begin{figure} [ht]
   \begin{center}
   \begin{tabular}{c}  
   \includegraphics[height=3.5cm]{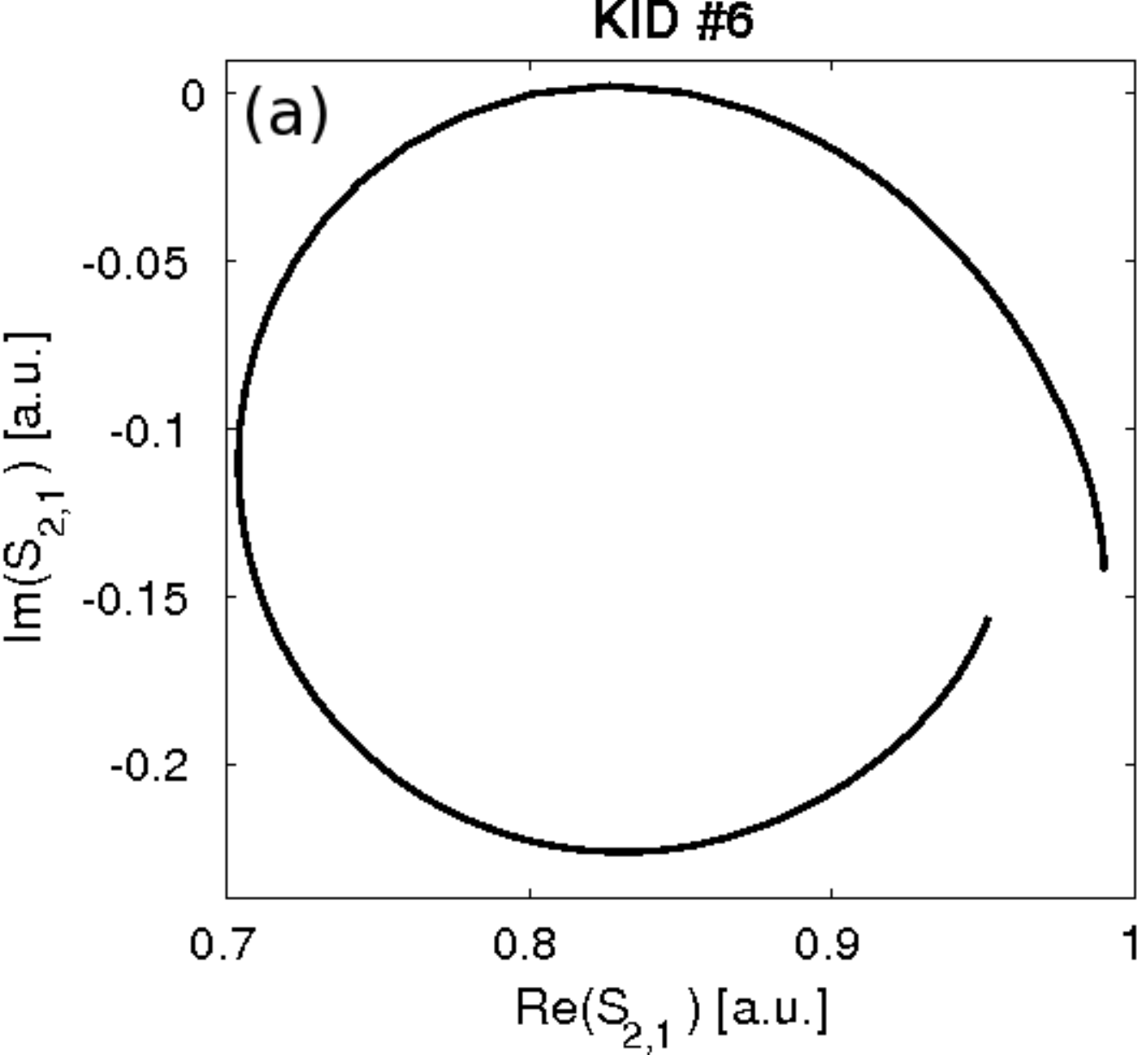}
   \includegraphics[height=3.5cm]{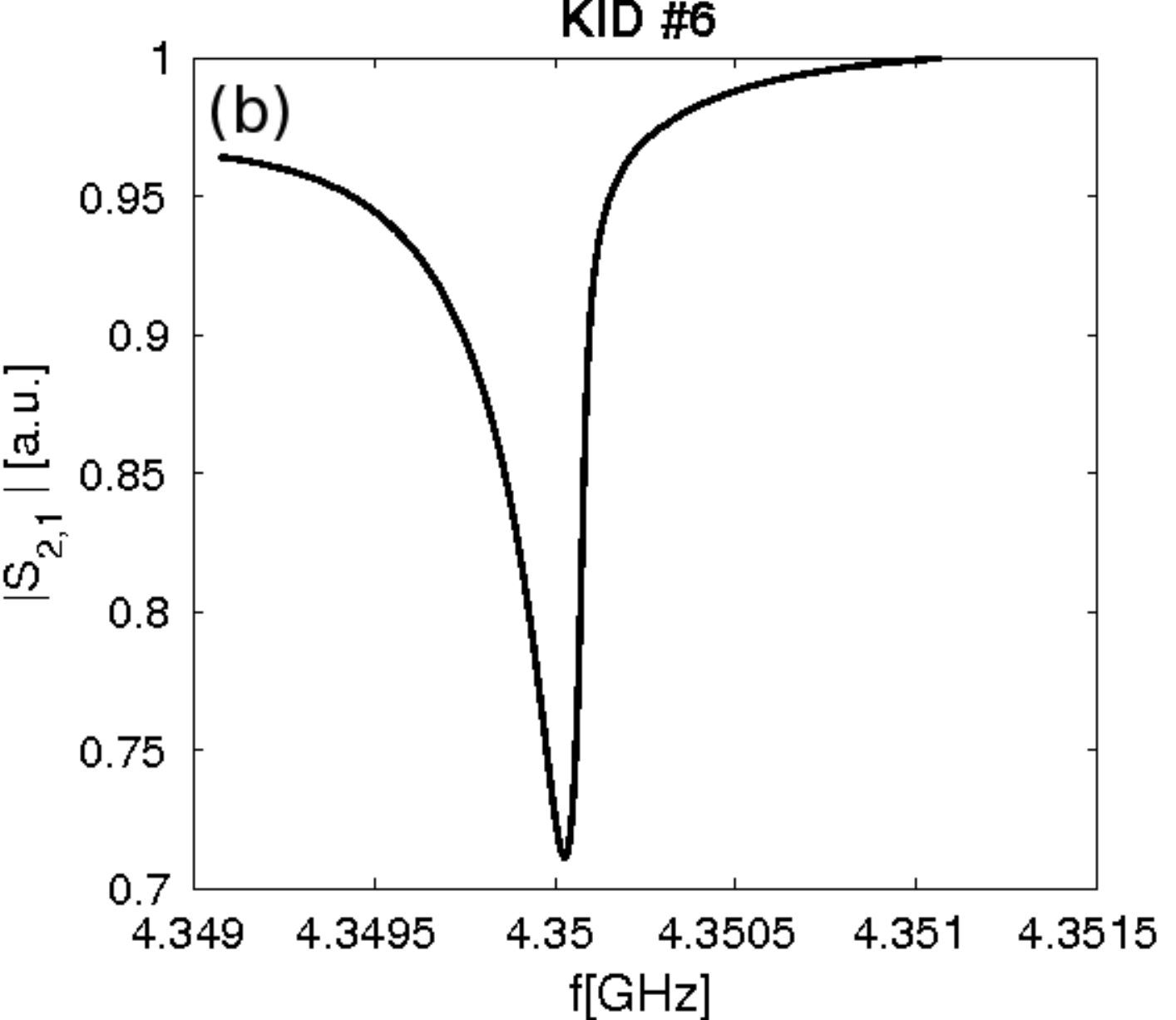}
   \includegraphics[height=3.5cm]{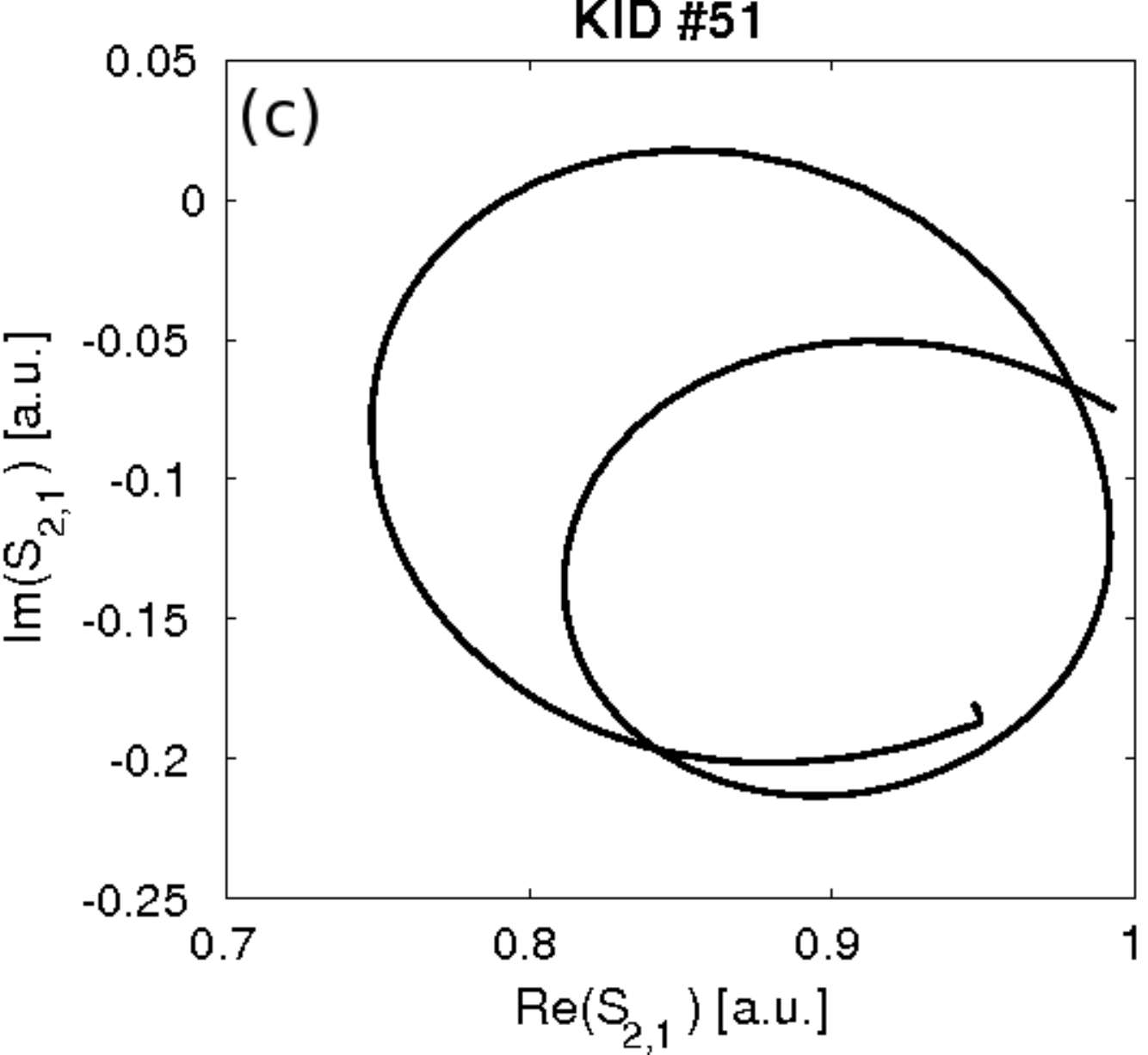}
   \includegraphics[height=3.5cm]{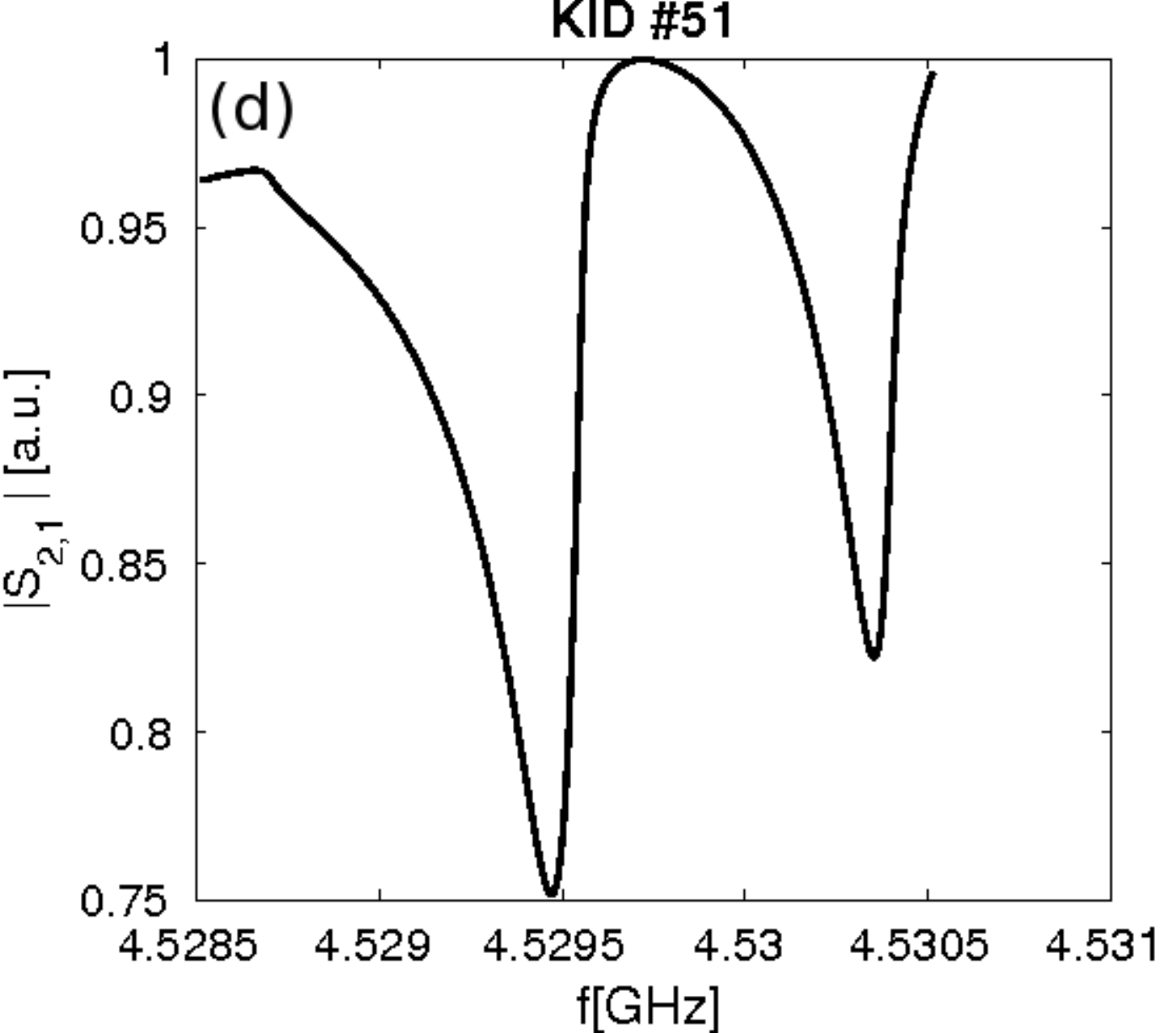}   
   \end{tabular}
   \end{center}
   \caption[example] 
   { \label{fig:S21} 
Transmission in the complex plane (a) and power of transmission versus the readout frequency (b) for an isolated KID. Transmission in the complex plane (c) and power of the transmission versus readout frequency (d) for two coupled KID.}
   \end{figure}

First, to recover the resonance frequency of each detector and analyse the level of crosstalk as a function of the separation in readout frequency, we measured the frequency sweep for each detector. In this way we obtained the complex transmission over a range in readout frequency of 2 MHz for all KIDs. It is possible to assess the presence of crosstalking detectors also by analysing the transmission. In particular, when a detector is isolated, its transmission is a circle in the complex plane (Fig. \ref{fig:S21}(a)) and a Lorentzian function in power (Fig. \ref{fig:S21}(b)). When crosstalk is present and the coupled detector is in the wavelength range scanned in the frequency sweep, two or more circles are visible in the complex plane (Fig. \ref{fig:S21}(c)), depending on the number of coupled KIDs, while the power of transmission is formed by two or more Lorentzian functions (Fig. \ref{fig:S21}(d)). \par

\begin{figure} [ht]
   \begin{center}
   \begin{tabular}{c}  
   \includegraphics[height=6cm]{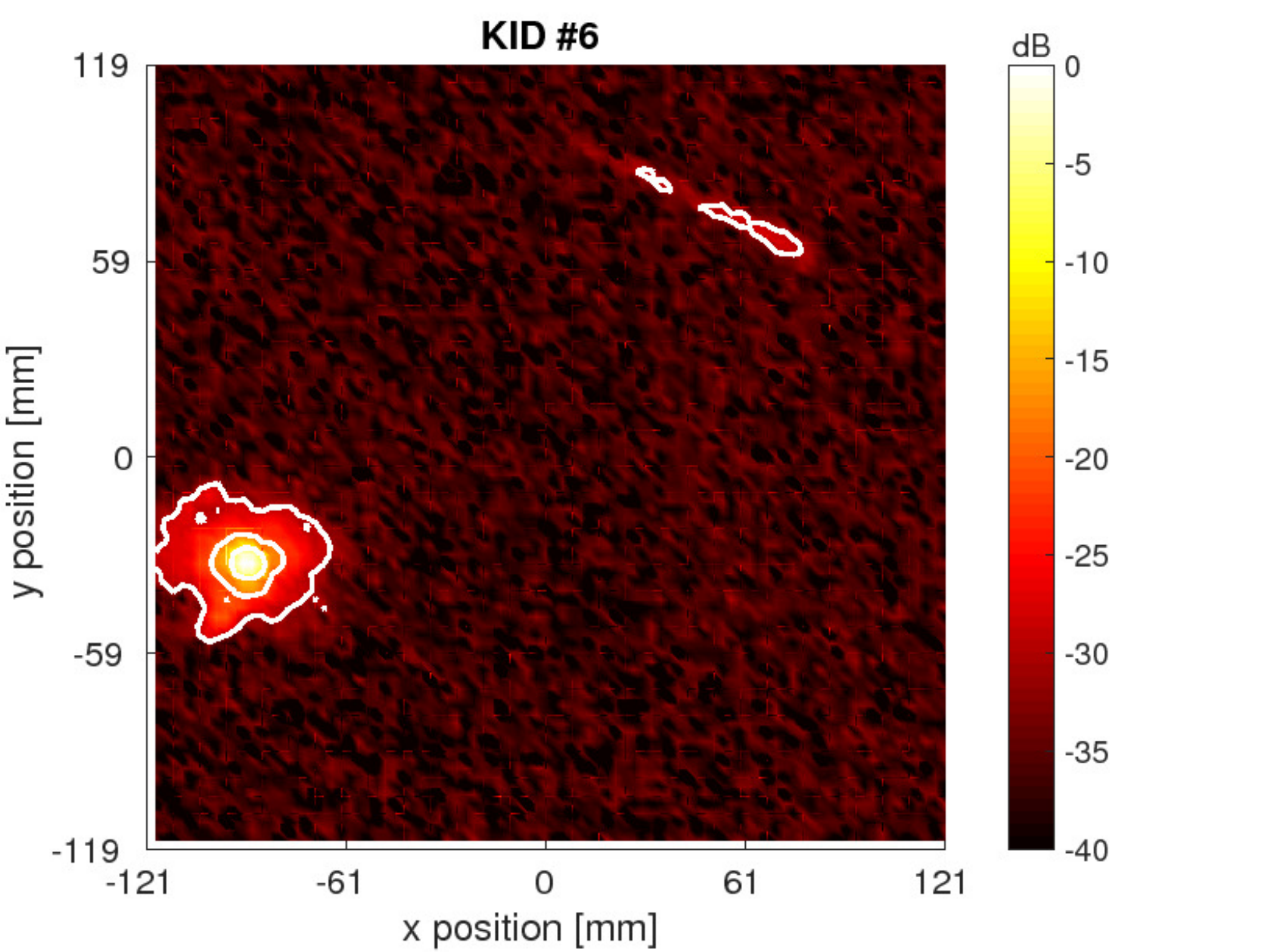}
   \includegraphics[height=6cm]{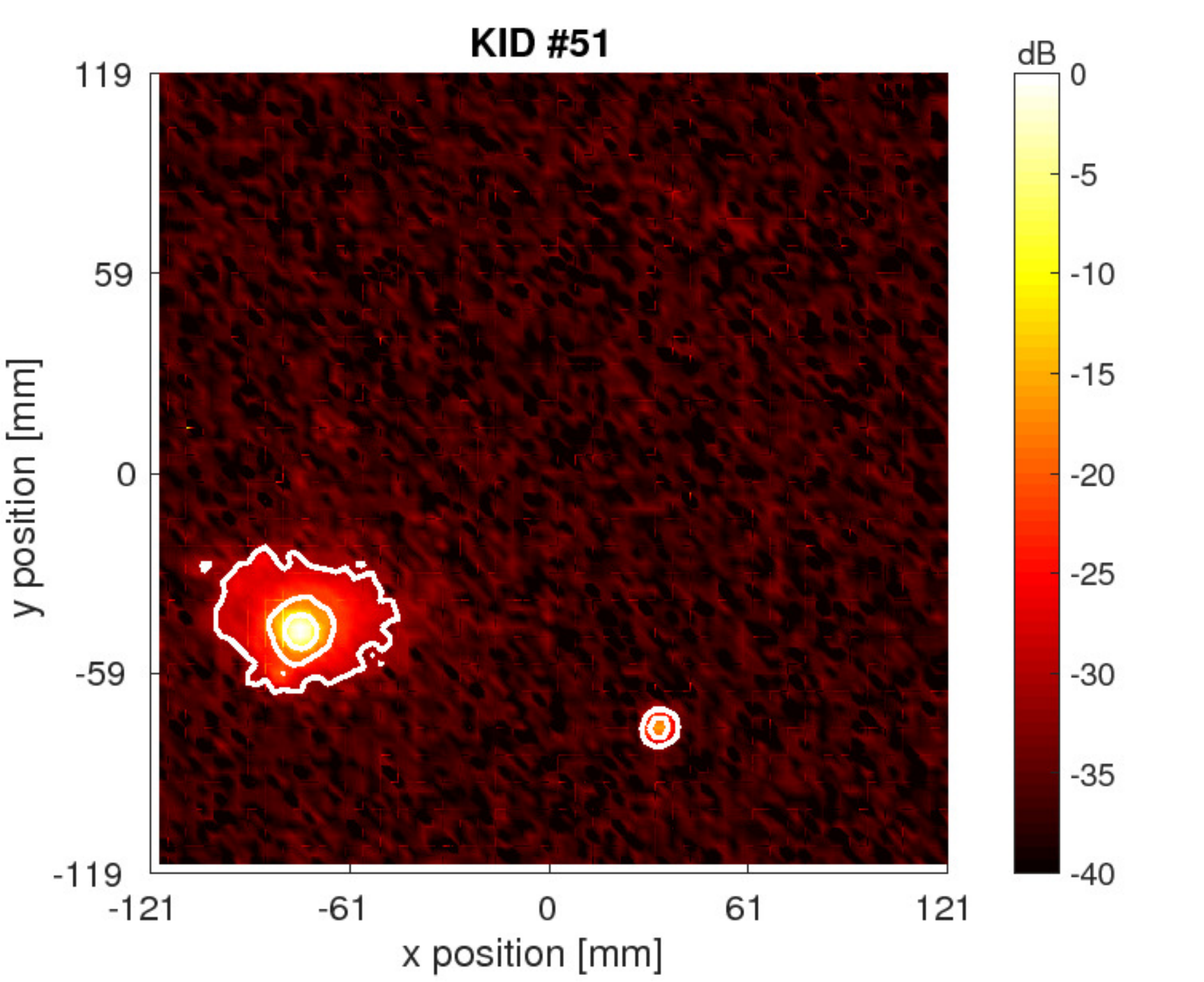}
   \end{tabular}
   \end{center}
   \caption[example] 
   { \label{fig:beammap} 
Two examples of beam maps. \textit{Left:} Beam map of a detector that is not cross talking. Only a PSF is present in this map. Some low level (-30 dB) ghosting from inside the cryostat is visible in the top right. This ghosting is not dealt with in the work, however in principle it can be corrected for. \textit{Right:}  Beam map of a detector that is coupled with another detector. Here a second response from the crosstalking KID is present. Contour lines are for -10, -20 and -30 dB from the maximum of the peak.}
   \end{figure}

 Second, we measured beam maps for all detectors in the array. We used the Groningen Beam-mapping facility that allows us to scan a hot source across the array by illuminating every detector once each time. The hot source is chopped at 80 Hz.  Further drifts in the total optical loading are removed by using linearisation via frequency sweep\cite{Bisigello2016}. This improves the linearity, but breaks down for very close KIDs where the frequency sweep is no longer described locally by a single Lorentzian. A 2 GHz subset of the array is read out at one time using the multiplexed readout presented in Ref.~\citenum{Ranrwijk2016}. This array is not designed to be read out with this readout, so 4 measurements at different local oscillator settings are required to get complete coverage of the entire array, with 400 to 200 pixels measured at the time. In the ideal case of absence of crosstalk, each beam map should contain a single image of the chopped source (Fig. \ref{fig:beammap} left), i.e. the PSF. In the case of crosstalk, there will be more than one peak in the beam map, corresponding to the expected response of the  KID of the map plus the response of crosstalking detectors (Fig. \ref{fig:beammap} right). Therefore, for each map we first identified every peak above -30 dB to recognise all detectors that are cross talking and, then, we measured the response of each KID in each map to obtain the level of crosstalk. \par
 \par

\subsection{PSF characterisation and cross-talk correction}
\begin{figure} [ht]
   \begin{center}
   \begin{tabular}{c}  
   \includegraphics[height=6cm]{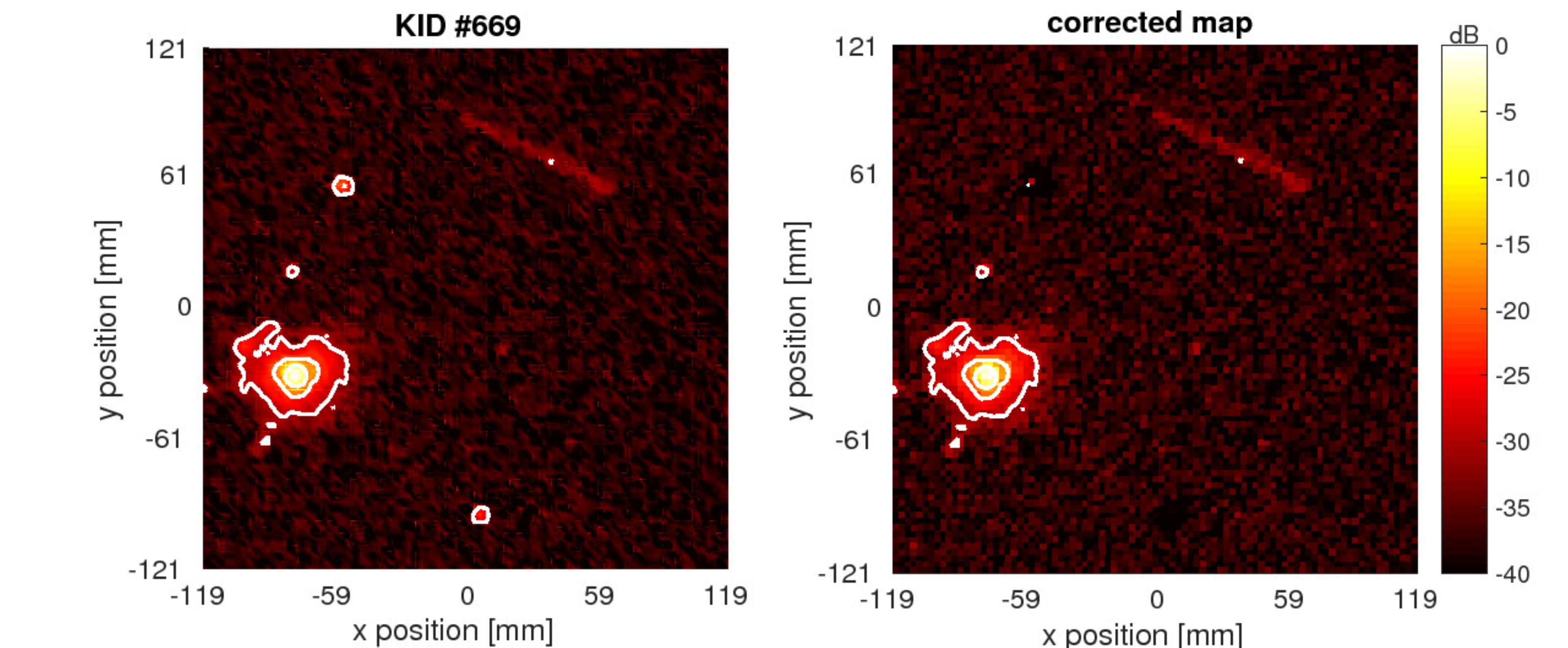}
   \end{tabular}
   \end{center}
   \caption[example] 
   { \label{fig:beammap_correction} 
\textit{Left:} This is an example of one original beam map where coupled KIDs are present, contour lines are for -10, -20 and -30 dB from the maximum of the peak. \textit{Right:} This is the same beam map, but after it has been corrected for crosstalk. The PSFs of each coupled KID have been subtracted from the original map, contour lines are for -10, -20 and -30 dB from the maximum of the peak.}
   \end{figure}

The response of a detector is the flux integral of the chopped signal in the image, therefore it is necessary to have a complete characterisation of the PSF to properly measure the response of each KID. With this intention, we considered a two-dimensional Gaussian beam to describe each PSF separately, in order to take into account for differences of the PSF through the array. After we derived the best fit for each PSF, we integrated the two-dimensional Gaussian to derive each response. Then, we subtracted the best fit two-dimensional Gaussian to each coupled detector to clean each map from the crosstalk (Fig. \ref{fig:beammap_correction}). 
\par
\begin{figure} [ht]
   \begin{center}
   \begin{tabular}{c}  
   \includegraphics[height=5cm]{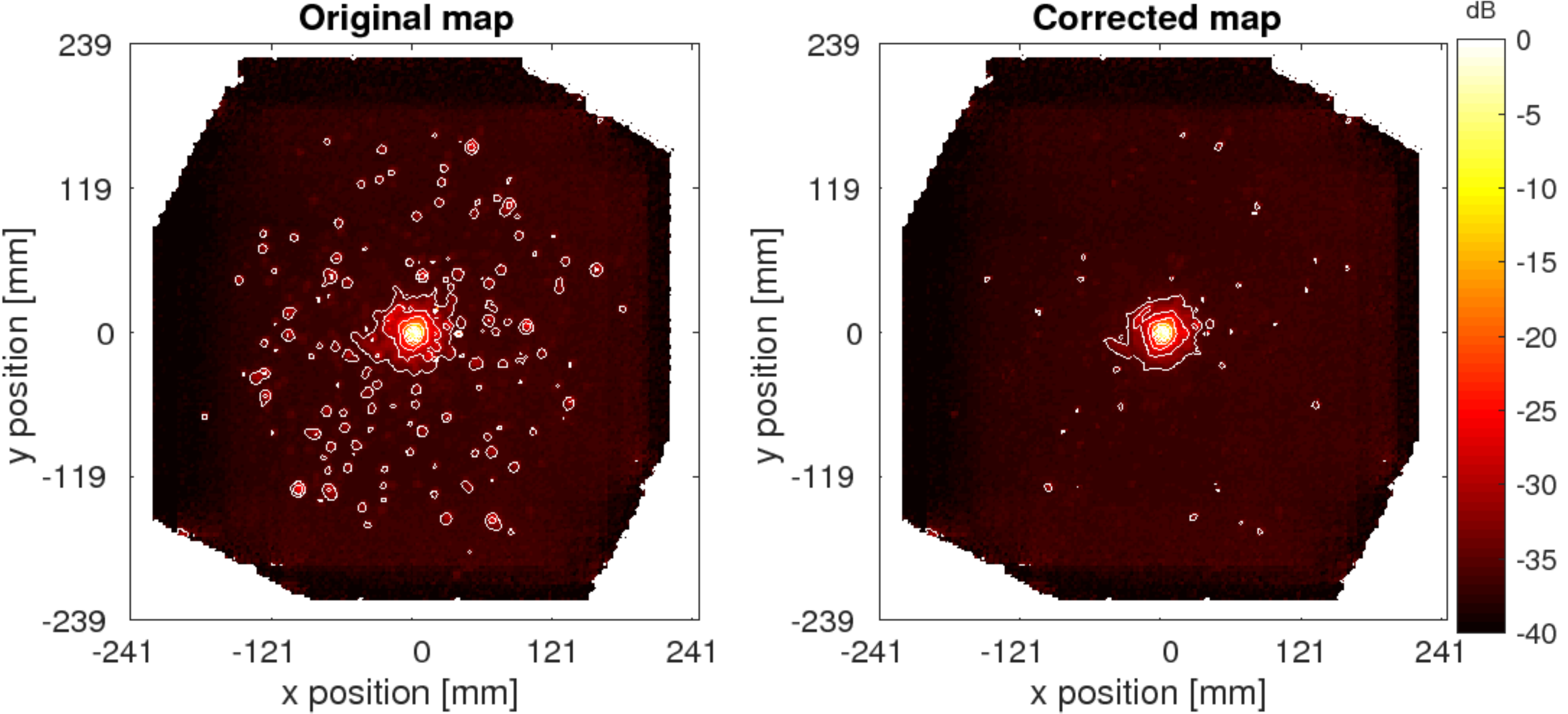}
   \includegraphics[height=5cm]{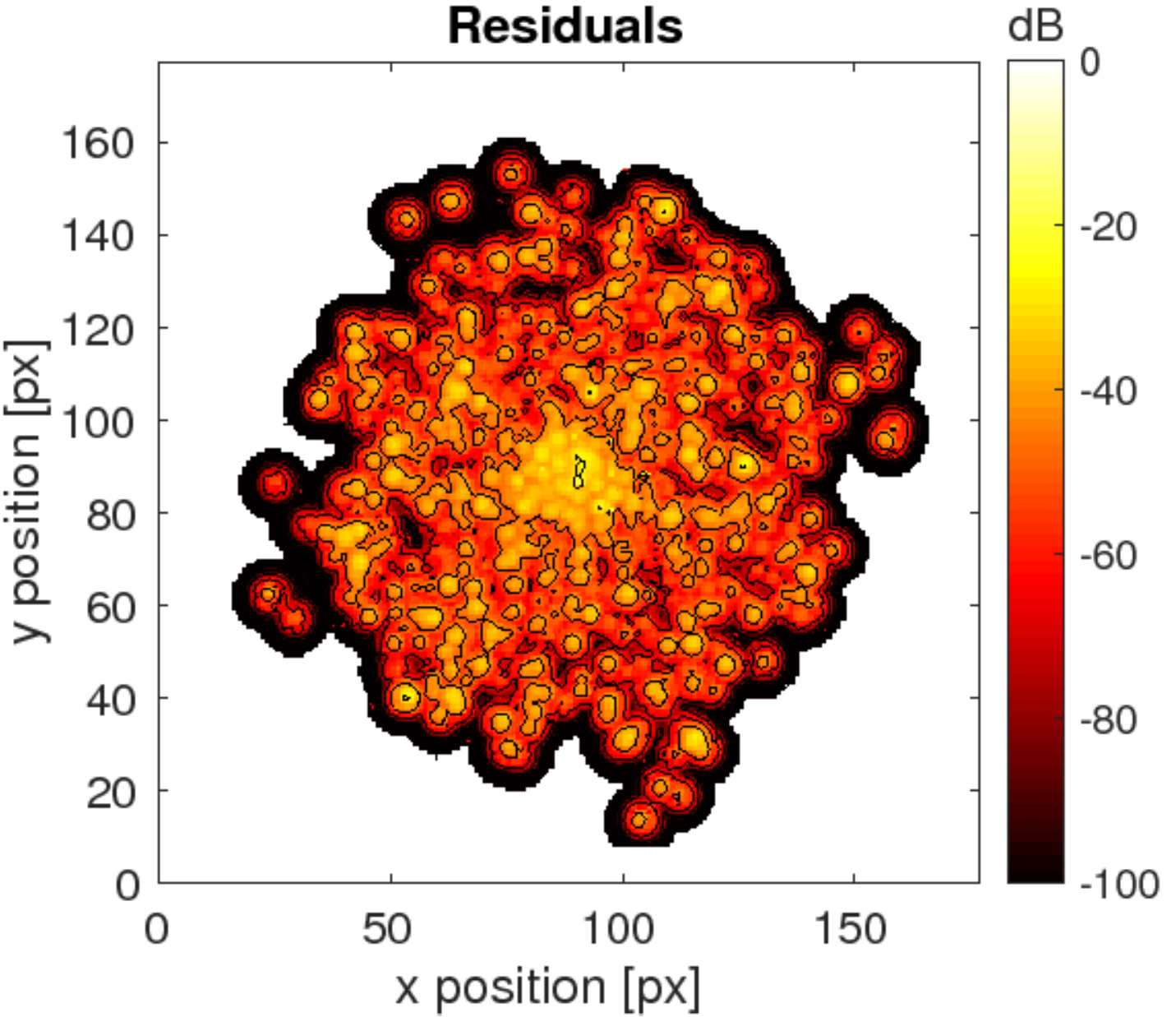}
   \end{tabular}
   \end{center}
   \caption[example] 
   { \label{fig:coadd}  Final map derived stacking all the original beam maps (\textit{left}) and all the corrected beam maps (\textit{center}). Contours lines represent differences of 5 dB. The right panel shows the residuals and contour lines indicate differences of 20 dB.
}
   \end{figure}

After we corrected every beam map for crosstalk above -30 dB from the maximum of the peak, we aligned all beam maps to the same beam centre and we stacked them together. In this way we obtained the image of the input chopped source. In Figure \ref{fig:coadd} it is shown the co-added map before and after the crosstalk correction, together with the residuals. Because the crosstalking KID are averaged out when creating the co-add map, their response is under -30 dB in the majority of the cases. For this reason, to evaluate the crosstalk correction applied in this work we also included the residuals, where all subtracted two-dimensional Gaussian are evident.

\section{Crosstalk level}
\label{sec:CT}
\begin{figure} [ht]
   \begin{center}
   \begin{tabular}{c}  
   \includegraphics[height=6cm]{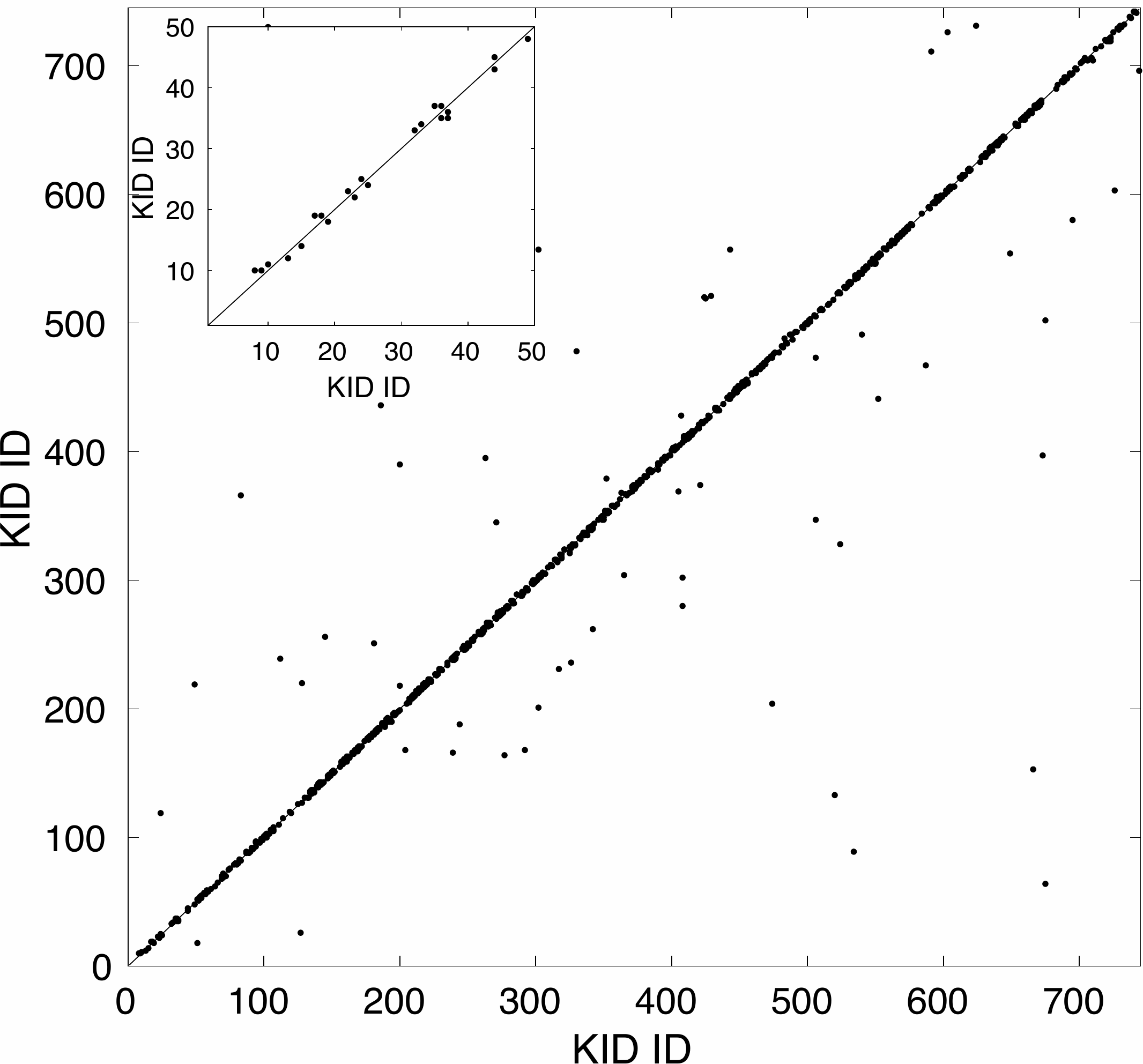}
   \end{tabular}
   \end{center}
   \caption[example] 
   { \label{fig:diagonal} 
Cross talking KIDs in our array. Both axis show the identifier numbers of detectors and the insert panel shows a zoom in of the first 50 KIDs. Since KIDs are sorted by their resonance frequencies, detectors with similar ID are close in readout frequency. Each point identify two KIDs that are cross talking and there are no point in the identity line, because a KID does not cross talk with itself. The majority of the points lies close to the one-to-one line, thus crosstalk is almost only due to KIDs close in readout frequency.}
   \end{figure}

Because we sorted our detectors by their resonance frequencies, KIDs with close identification numbers are also close in readout frequency. Around 15$\%$ (136 out of 880) of all detectors were not identified during measurements, therefore we did not measure neither beam patterns nor frequency sweeps for these KIDs. Because we could not derive their resonance frequencies, we did not include them in the crosstalk analysis. However, they have been removed in the beam map where they can still been observed. By comparing the numbers of KIDs that are cross talking, it is possible to derive the nature of the crosstalk. It is evident from figure \ref{fig:diagonal} that most coupled detectors are close in readout frequencies. Those few cases that are distant in readout frequency are likely caused by reflections or noise in determining the crosstalk, e.g. missidentification and missfited KIDs. Therefore, this support our assertion earlier that the main cause of crosstalk in this array is the overlap of readout frequencies and the level of crosstalk decreases with the increase of the separation in readout frequencies. About 28$\%$ of detectors in this array are isolated and are not coupled with any other detector above -30 dB from the maximum. On the other hand, $\sim$48$\%$ of all KIDs are coupled with another detector, $\sim$16$\%$ with other two KIDs and $\sim$7$\%$ with other three KIDs.
     \par
\begin{figure} [ht]
   \begin{center}
   \begin{tabular}{c}  
   \includegraphics[height=8cm]{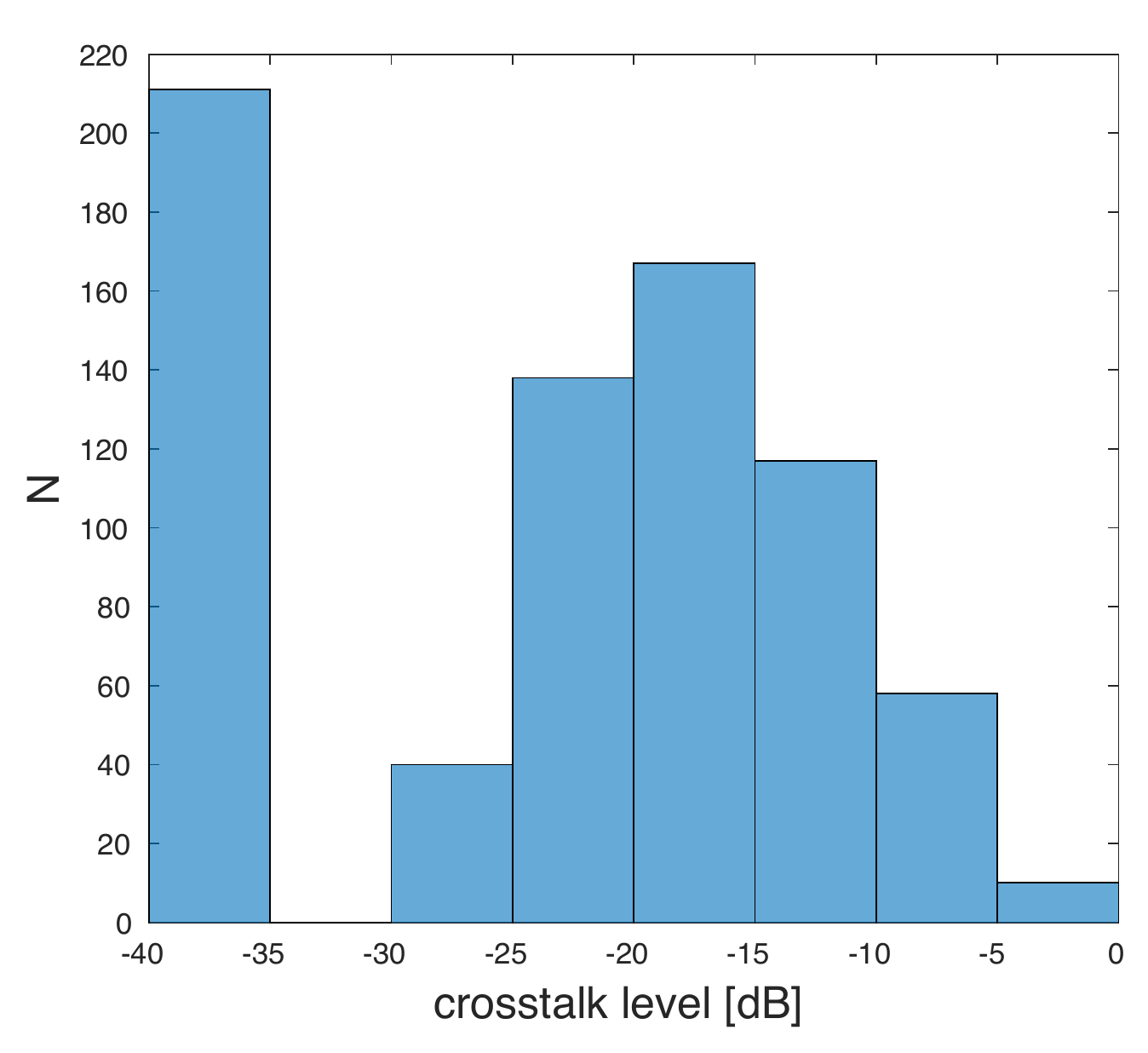}
   \end{tabular}
   \end{center}
   \caption[example] 
   { \label{fig:histogram} 
Level of crosstalk present in the array. We assigned to isolated KIDs a level of crosstalk of -40 dB.}
   \end{figure}

By using the response derived from the two-dimensional gaussian fitting it is possible to calculate the level of crosstalk for each coupled KID. The level of crosstalk between two KIDs given by:\begin{equation}
\label{eq:fov}
K_{1,2} = \frac{R_{2}}{R_{1}} \, ,
\end{equation}
where R$_{1}$ is the response of KID 1 that was illuminated, R$_{2}$ is the response of the coupled KID, called KID 2, as a result of the signal measured by KID 1. Thus K$_{1,2}$ is the level of crosstalk of KID 2 due to KID 1. It is worth mention here that, in general, K$_{1,2}\neq$ K$_{2,1}$. In this way we measured the level of crosstalk for the array, as it is shown in figure \ref{fig:histogram}, where we considered the maximum level of crosstalk for each KID and we assigned a crosstalk level of -40 dB to isolated detectors. We found that around 48$\%$ of all detectors are at least coupled with an other detector with K$_{1,2}>-20$ dB.
\par
\begin{figure} [ht]
   \begin{center}
   \begin{tabular}{c}  
   \includegraphics[height=8cm]{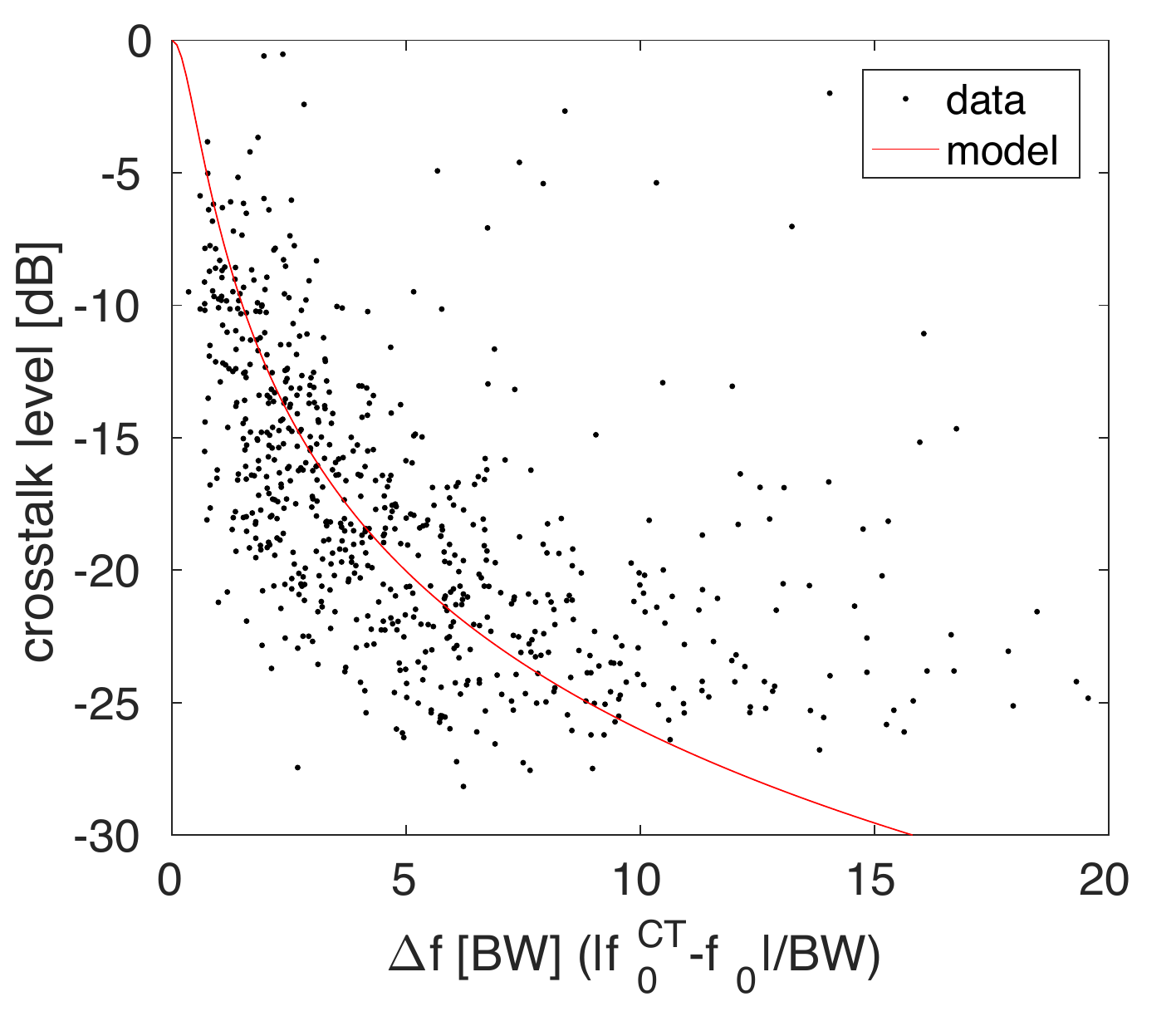}
   \end{tabular}
   \end{center}
   \caption[example] 
   { \label{fig:CTlevel_Df} Crosstalk level versus readout frequency separations in units of band widths for all coupled KIDs of the array (black points). The red dashed line is the theoretical description given by equation \ref{eq:CTlevel1}.
}
   \end{figure}

In addition, we derived the resonance frequency and Q-factor of each KID from the frequency sweep, in order to analyse the dependence of the crosstalk level with the readout frequency separation of the detectors. In particular, we fitted a Lorenztian function to the power of the transmission measured in the frequency sweep\cite{Mazin2005}:
\begin{equation}
\label{eq:S21Fit}
S_{2,1}^2=1-\frac{(1-S^2)}{1+(\frac{2Q(f-f_0)}{f})^2} ,
\end{equation}
where S$_{2,1}$ is the transmission, S is the minimum of the dip transmission, f$_0$ is the resonance frequency, Q is the quality factor of the considered KID and f the analysed readout frequencies. From this fit we derived both the resonance frequency and the Q-factor of all KIDs, as well as the bandwidth.
We then compared the level of crosstalk with the readout frequency separation in band widths of the crosstalking KIDs (Fig. \ref{fig:CTlevel_Df}). In general, the level of crosstalk decreases with the increasing readout frequency distance of the crosstalking KIDs, with a final plateau probably due to the noise level of the beam maps.\par
Additionally, we described the relation between the crosstalk level and readout frequency separation, first with a simple model and then with a more complex one. We started from the fact that the response of a KID is a Lorentzian in power. From equation \ref{eq:S21Fit} and defining X$\equiv Q\frac{f-f_0}{f_0}$, we obtained that the power and the phase of the transmission can be written as:
\begin{equation}
\label{eq:S21}
|S_{2,1}|^2=1-\frac{1-S^2}{1+4X^2}\\, 
\end{equation}
\begin{equation}
\label{eq:theta}
\tan(\theta)=\frac{4X}{X^2-1},
\end{equation}
where S$_{2,1}$ is the transmission and $\theta$ is the phase, S is the minimum of the dip transmission, f$_0$ the resonance frequency and Q the quality factor of the KID. Then, from eq. \ref{eq:theta} we can derive that:
\begin{equation}
\label{eq:dx}
 \frac{\delta\theta}{\delta X}=\frac{4(4X^2+1)}{16X^2+(4X^2-1)^2},
\end{equation}
If we assume that the two coupled KIDs have the same dip depth and Q-factor, their complex transmission will be the same. Therefore, a signal will produce the same phase shift in both detectors. Moreover, we can consider that $\Delta\theta=\Delta X \cdot \delta\theta/\delta X$ and, from equation \ref{eq:dx}, we can derive that $\theta=$4X in the limit for f$\to$ f$_{0}$, assuming that system effects are calibrated out. Therefore, the crosstalk level between two KIDs can be expressed as:
\begin{equation}
\label{eq:CTlevel1}
K_{1,2} = \frac{\Delta X_{2}}{\Delta X_{1}}=\frac{1}{4}\frac{\delta\theta}{\delta X}=\frac{(4X^2+1)}{16X^2+(4X^2-1)^2}
\end{equation}
By considering that dip depths and Q-factors are the same for the all KIDs, it is possible to describe the general relation between crosstalk level and separation in readout frequencies (Fig. \ref{fig:CTlevel_Df}). \par
\begin{figure} [ht]
   \begin{center}
   \begin{tabular}{c}  
   \includegraphics[height=8cm]{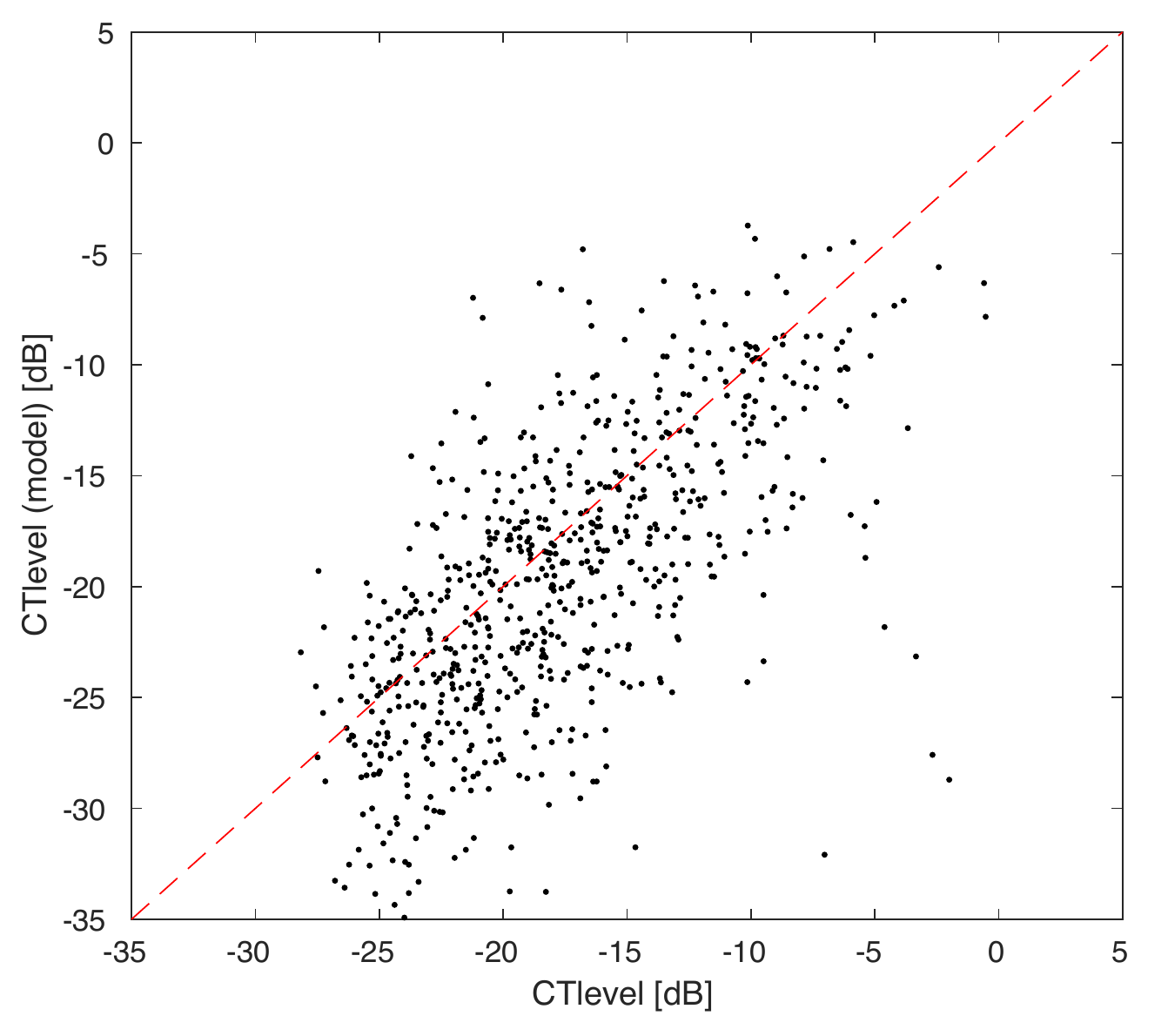}
   \end{tabular}
   \end{center}
   \caption[example] 
   { \label{fig:CTlevel_Df2} Comparison between the measured crosstalk level (x axis) and the theoretical one (y axis) given by equation \ref{eq:CTlevel2}. The dashed line rappresent the one-to-one relation.
}
   \end{figure}

In order to predict the crosstalk level for each KID more precisely, we include in equation \ref{eq:CTlevel1} both the Q-factor and the dip depth of each KID. The resulting formula is:
\begin{equation}
\label{eq:CTlevel2}
K_{1,2} =\frac{\Delta X_{2}}{\Delta X_{1}}=\frac{1}{4}\frac{Q_1}{Q_2}\frac{(1-S_{2})}{(1-S_{1})}\frac{\delta\theta}{\delta X}
\end{equation}
where Q$_{i}$ and S$_{i}$ are the Q-factor and the minimum of the dip transmission for the KID i, respectively. The comparison between the measured crosstalk level and the theoretical one is shown in figure \ref{fig:CTlevel_Df2}. 
The root mean square of the difference between the modelled crosstalk and the measured one is $\sim$5 dB. The scatter could be due to missidentified (e.g. missing) and missfitted KIDs. In addition, the KID shape has strong readout power dependence\cite{deVisser2014} so is not described by only a single Lorentzian fit to the frequency sweep, particularly off-resonance. To conclude, a model that takes into account more parameters is necessary to describe the crosstalk level of the full array, but this model is still valid as first order approximation.

\section{Summary}
\label{sec:end}
In this work we characterised the level of crosstalk of an MKID array in order to correct images for crosstalk a posteriori.
We measured beam maps for all KIDs in the array and we described the PSF as a two-dimensional Gaussian, for each KID with response above -30dB from the maximum. We subtracted the best-fit PSF of all coupled KID from the original beam map in order to remove the crosstalk present. Following this procedure, it is possible to correct astronomical images for crosstalk a posteriori. Analysing the level of crosstalk present in the array, we derived that about 72$\%$ of KIDs in the array crosstalk above -30 dB level while $\sim$48$\%$ crosstalk above -20 dB level. In the full array, 48$\%$ of all KIDs are coupled to another detector, 16$\%$ are coupled to other two detectors and 7$\%$ are coupled to other three KIDs. \par
We estimated the resonance frequency and the quality factor of each KID by measuring the frequency sweep and describing the power of the transmission as a Lorenztian function. By using these parameters we derived a model by assuming that all KIDs have the same dip depth of the transmission and Q-factor. This model describes the expected general level of crosstalk as a function of the readout frequency separation of the detectors. This shows, both experimentally and using a simple model, that the rule of thumb is that KID-KID separation higher than 10 KID bandwidth corresponds to $\sim$-25 dB crosstalk. This can be taken as design criterion, as required, and as a way to estimate crosstalk for future arrays.

\acknowledgments 
The authors thank the team of the SPACEKIDS project for the collaboration.This project was supported by ERC starting grant ERC-2009-StG Grant 240602 TFPA and Netherlands Research School for Astronomy (NOVA). 

\bibliography{SPIE2016.bbl} 
\bibliographystyle{spiebib.bst} 

\end{document}